# Effect of milling on dispersion of graphene nanosheet reinforcement in different morphology copper powder matrix

N. Vijay Ponraj[a], S.C. Vettivel[b], A. Azhagurajan[c], X. Sahaya shajan[d], P.Y. Nabhiraj[e], T. Theivasanthi[f], P. Selvakumar[a], A. Haiter Lenin[g]

a. Department of Mechanical Engineering, PSN College of Engineering and Technology, Tirunelveli, Tamilnadu, India
b. Department of Mechanical Engineering, Chandigarh College of Engineering & Technology (CCET) – Degree Wing, Chandigarh, 160017, India
c. Department of Mechanical Engineering, Mepco Schlenk Engineering College, Sivakasi, Tamilnadu, India
d. Centre for scientific and applied research, PSN College of Engineering and Technology, Tirunelveli, Tamilnadu, India
e. Variable Energy Cyclotron Centre, Bidhannagar, Kolkata, 700064, Tamilnadu, India
f. International Research center, Kalasalingam Academy of Research and Education (Deemed University), Krishnankoil, 626126, Tamilnadu, India
g. Department of Mechanical Engineering, WOLLO University, KIOT, Ethiopia

Emails: ttheivasanthi@gmail.com , vijayponrajmechauto@gmail.com
ORCID: https://orcid.org/0000-0002-2280-9316

___________________________________________________________________________

**Abstract:** The effective dispersion of Graphene Nanosheet (GNS) as reinforcement was studied with various morphology of Copper (Cu) powder as matrix. High energy milling was used for modifying the morphology of the Cu powders. The Cu powder with Spherical and dentric shape was used in this study. Using high energy milling both the powders were milled for 8 h and 16 h respectively. The morphology of Cu powders was altered as flake shape upon milling. GNS (2 wt. %) was added as reinforcement uniformly with various morphology Cu powder used as matrix. The effect of size and shape of Cu as matrix on Cu/GNS composite properties was comprehensively studied. The Cu/GNS composite was prepared using powder metallurgy technique. Using the different shape and size of the Cu as matrix the interface of GNS has been studied. The hardness properties of Cu/GNS composite are evaluated. The effect of GNS interface in Cu particles has a significant influence on mechanical properties of composites. The hardness of Cu matrix composite has improved up to 20% compared to that of pure Cu. Thus morphology of Cu has the ability to improve the mechanical properties with GNS reinforcement.

**Keywords:** Copper, Graphene nanosheet, High energy milling, SEM, Metal matrix composites

___________________________________________________________________________

## 1. Introduction

Composite is the combination of two or more materials which are having different phases and the properties superior to the base material. Metal Matrix Composite (MMC) applications are more in bimetallic metals in the field of mechanical, aerospace, automotive and electrical industries [1]. High strength, high electrical and thermal conductivity with ductility make MMC peculiar than bimetallic metals. Copper (Cu) reinforced with nano carbon materials earns more attention due to its enhanced mechanical properties without affecting its electrical and thermal properties. Carbon materials such as carbon nanotubes, graphene nanosheet, diamond, graphite are highly used as reinforcement with Cu.

Graphene nanosheet (GNS) has received considerable interest in research due to their extraordinary high elastic modulus (1 TPa) and fracture strength (125 GPa) [2] as well as with high flexibility [3]. Other than their exceptional mechanical properties, GNS have extremely high surface area [4], excellent chemical stability [5] as well as higher thermal [6] and electrical properties [7]. All these characteristics have made GNS outstanding reinforcement materials for developing advanced nanocomposites for cryogenic, nuclear, space applications [8]. Graphene dispersed phase has great potential applications in aerospace or electronic fields as advanced composite materials [9].

Considerable researches have performed on GNS reinforced polymer matrix composites with significant enhancement in mechanical properties compared to those of base metal [9–11]. However, few papers have been done in preparation and analysing mechanical properties of metal/GNS reinforced composites [12–16]. Due to drawbacks suchs as poor wetting behaviour, weak interfacial bonding to metal matrix materials, agglomeration among themselves with van-der-Waals force [17], inhomogeneous distribution of GNS in the matrices and degraded thermal stability at high processing temperature make GNS as complicated reinforcement with metal [18].

Several production methods have been used for fabricating Cu/GNS composite such as powder metallurgy, accumulative roll bonding, ball milling, electro deposition. Among the above methods powder metallurgy assisted with ultra sonication technique shows high uniform dispersion of GNS and excellent mechanical properties [12]. High energy milling is an important method for producing composites materials with various morphology, improved mechanical properties etc. Also it is an inexpensive and fast method.

Previously, Ponraj et al. [35] successfully fabricated Cu/GNS composite using powder metallurgy technique with Polyvinyl Alcohol (PVA) treatment followed by sintering shows 10% enhancement of compressive strength compared to pure Cu matrix with 0.2 Wt.% of GNS [13]. Moreover, GNS reinforcements with other process have so far been reported poor efficiency [18] due to agglomeration among themselves with Van-der-Waals force, poor wetting behaviour or weak interfacial bonding to matrix materials.

Graphene has higher strength than other materials [19]. Hence, it can be a suitable candidate to be added as a filler material in MMC to improve the mechanical properties [20]. After high energy milling, morphology and mechanical properties are changed. The spherical and dendritic structures of Cu powder are converted into micro layer structure and hardness is improved. Both micro layer structure and hardness improvement are the results of the penetration of two dimensional GNS into Cu powder during milling [21].

In order to explore the full potential of GNS reinforced Cu, in this work; we authors take effort to enhance the bonding between Cu and GNS with surface modification technique. Since GNS is in nano-level and modifying its surface is highly difficult. On other hand Cu particles are used in micron level. Using high energy milling the Cu powders surface modified. Also the literature confirms that nobody have studied the uniform distribution of GNS in the Cu matrix. This paper is going to deal with the uniformity in dispersion of GNS with various morphology Cu powders and its mechanical properties.

## 2. Experimental

### 2.1. Materials

Sodium nitrate, NaNo3 (99.99%), Cu spherical powder supplied from Sigma-Aldrich, Graphite, Sulphuric acid (H2So4), Hydrogen Peroxide, Potassium Permanganate, Hydrochloric acid supplied from Rankem chemicals. All chemical reagents were analytical grade and distillation deionized water.

### 2.2. Preparation of Cu/GNS composite

High energy milling was done on Cu powder for size reduction using planetary micro mill (pulverisette 7 from fristch Gmbh) with tungsten carbide balls in Tungsten carbide vial. The ball to powder ratio used is 10:1. In our previous work, the Cu/GNS composite fabricated using powder metallurgy technique exhibit 9% enhancement in mechanical properties than pure Cu. PVA is a highly hydrophilic material [22, 23]. Cu powders treated with 3 wt % of PVA for creating a hydrophilic effect on the surface [23]. Graphene Oxide (GO) is prepared by modified Hummers method [24]. Aqueous GO is added to the enhanced hydrophilic Cu powder and then stirred with a mechanical stirrer. Prepared Cu/GO composite powders heated at 650 °C at a heating rate of 40 °C min$^{-1}$ and kept in the flow of nitrogen for 2 h. Upon heating the GO reduces to GNS and PVA decays. Cu/GNS composites fabricated by using suitable die sets at high load (1 GPa) the specimen fabricated at 12mm diameter and 12mm long. The cold compacted sample sintered at 750 °C using muffle furnace kept at flowing nitrogen gas. After sintering, sintered composites were cooled inside the furnace at room temperature.

## 3. Results and discussions

### 3.1. Evaluation of mixing

Fig. 1(a) shows the Scanning Electron Microscope (SEM) image of pure Cu powder as received. Fig. 1(b) shows the SEM image of dendritic structure Cu powders. Fig. 1(c & d) shows the SEM and Transmission Electron Microscope (TEM) image of GO prepared using hummer's method. The morphology of powder is characterized by spherical in size with an average size of 45 μm. Almost all the particles are nearly even in size and shape also revealing rough surface. Fig. 1(b) shows dendritic morphology powder with an average size of 70 μm. Fig. 1(c) indicates the morphology of GO powder. The thickness of the sheet is found to be approximately 5–30 nm in size and height of the sheet is up to 0.8 nm [13]. Poor intermolecular adhesion is due to van der Walls forces. GO has randomly stacked, associated and confused multilayered nanosheet structure with wrinkles. Enough voids are present to facilitate the formation of nanosheets under sonication.

Fig. 2(a–d) shows the Cu powder with different morphology after reinforced with GNS. Reinforcing GNS with Cu is donethrough PVA assisted technique and thermal reduction method. Initially, GO is sonicated for 2 h to exfoliate graphene layers sheet by sheet. On another side, PVA is treated with Cu particles to create wetting on the surface of Cu particles. Then the exfoliated GO is added drop by drop in PVA treated Cu under mechanical stirrer at 400 rpm. By heating, the mixed Cu/GO slurry at 650 °C the PVA and oxides are removed. Fig. 2(a) shows the Cu/GNS composite powder, GNS reinforced with Cu powder as on received state; it shows uniform dispersion of GNS. Majority of Cu powders are covered with GNS.

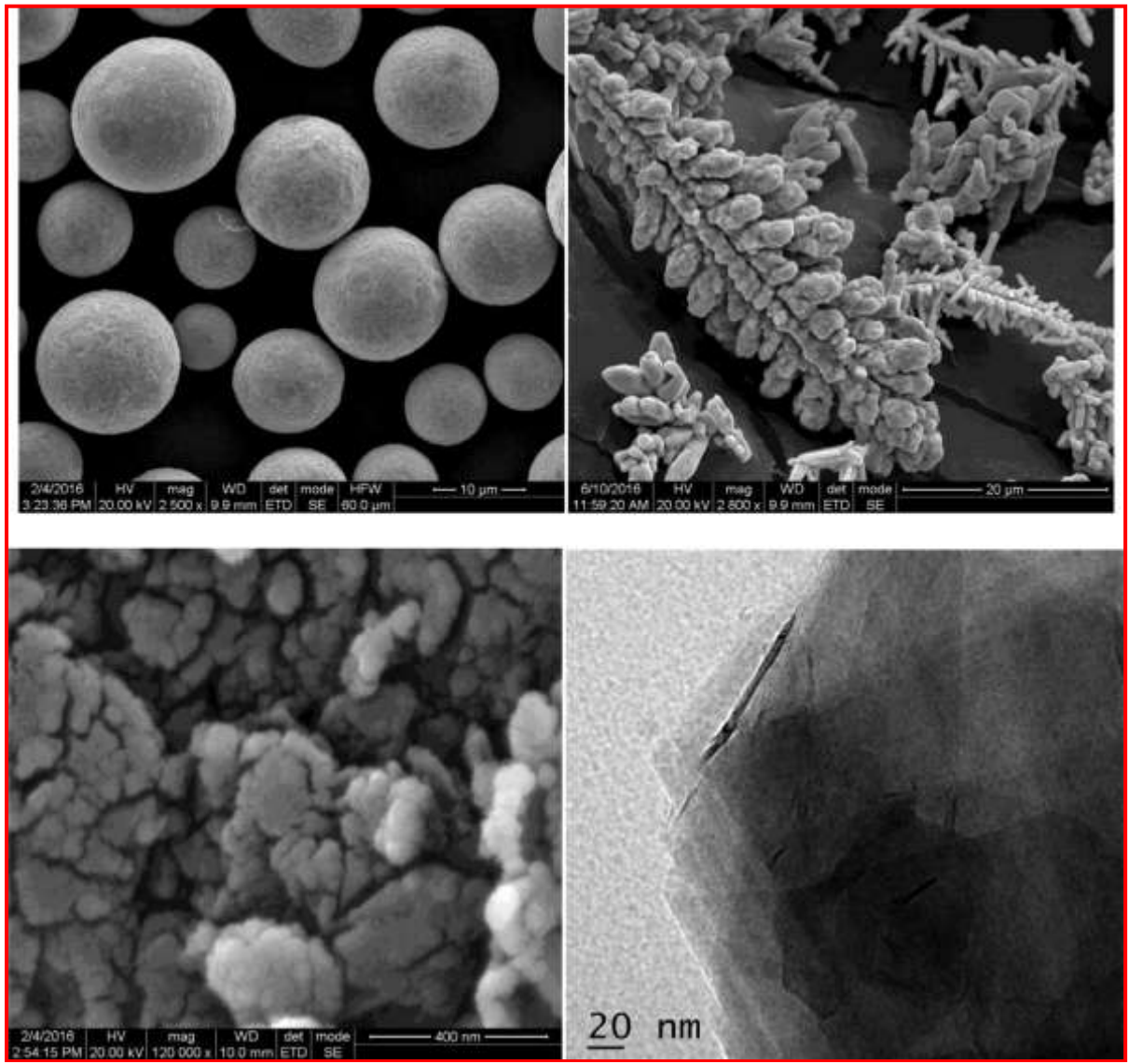

**Fig. 1.** (a) SEM image of pure spherical Cu powder (b) SEM image of pure dendritic Cu powder (c) SEM image of prepared GO (d) TEM image of prepared GO.

The planetary ball mill is employed for modifying the Cu powder size by strong centrifugal force to increase high energy milling inside the tungsten carbide vial. Milling process involves continuous impact, welding, fracturing and re-welding of Cu powders. Fig. 2(b) shows (Milling for 8 h) cold-welding of Cu powders which enhance the size to enormous metal particles by reducing thickness to 5 μm. By reducing the particle size of metal particles, the possibility of dispersing GNS in metal powders is increased. In Fig. 2(c) 1 μm thick dendritic Cu particles after 8 h milling reinforced with GNS shows good dispersion of GNS but major area have not covered with reinforcement [25–30]. Fig. 2(d) shows nanostructured Cu powder of 200 nm size after 16 h milling, which results in homogenous dispersion of GNS, almost every area of Cu particles is covered with GNS.

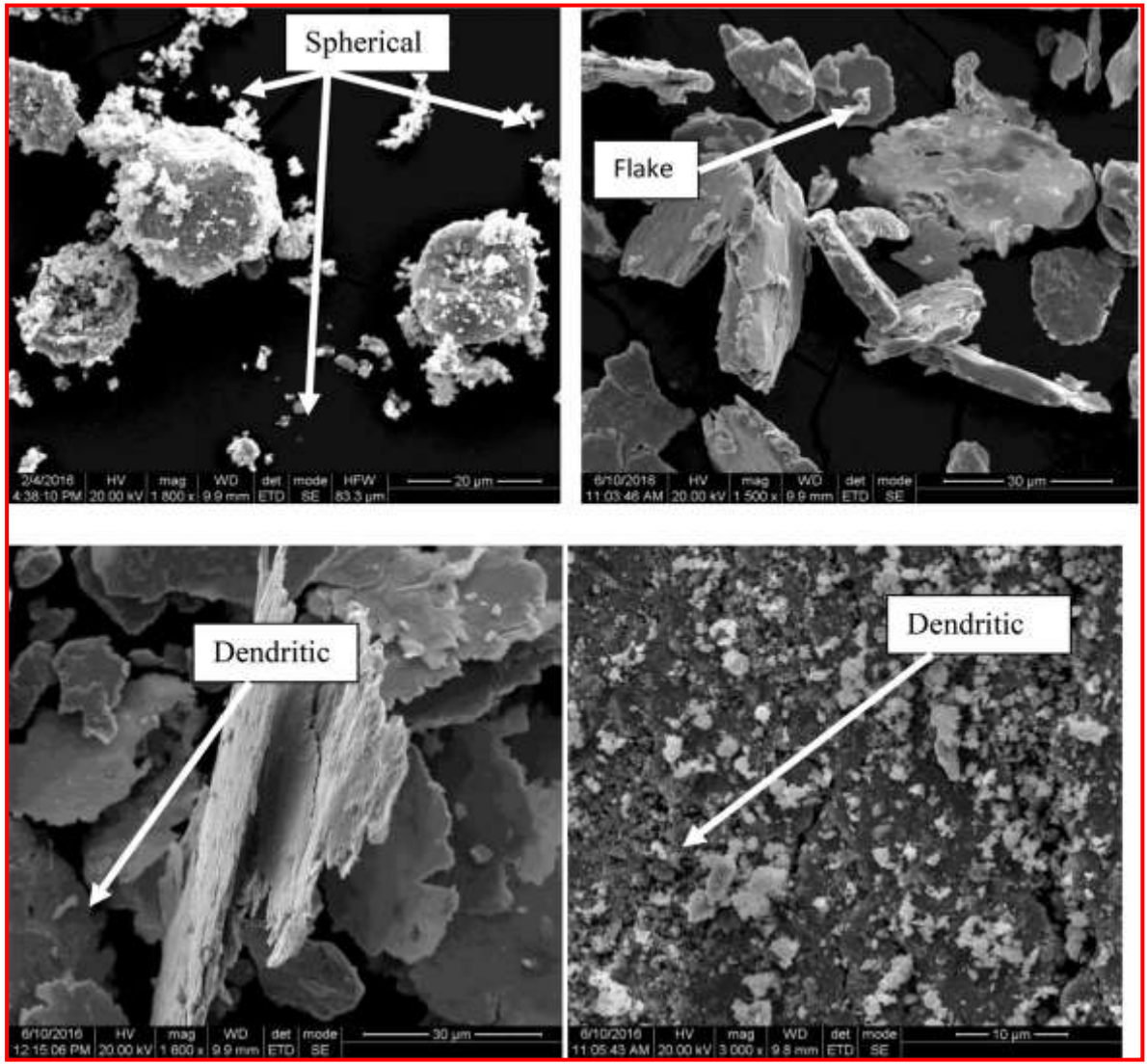

**Fig. 2.** shows the GNS reinforced with (a) spherical powder (b) spherical powder after 8 h milling (c) dendritic powder after 8 h milling (d) dendritic powder after 16 h milling

Fig. 2(a–d) shows the homogeneously distributed GNS among the Cu powder through this method. Good interfacial bonding was shown. Homogenous distribution of GNS observed in all shapes of Cu. This homogenous distribution achieved by exfoliated GO added drop by drop under mechanical stirring, makes the distribution of GNS uniformly over the entire area of Cu powder. PVA treated with Cu particles to increase the wet-ability of Cu powder [31], PVA produces strong hydrogen bonding between Cu and GO [32]. PVA prevents agglomeration of GNS and stabilize the separated nano-sheets under heating [33]. Since PVA is treated with the matrix, it enhances the GO tends to tangle and agglomerate into clusters, due to bonding under weak vander Waals forces. Upon heating at 650 °C, the PVA decays and GO is changed to GNS by removal of oxides through heating at the flow of inert atmosphere. Good interfacial bonding proves that this method is suitable for reinforcing GNS with any matrix of different size and shape. Fig. 3(a–h) shows the deep magnification performed using SEM–Energy Dispersive Spectrum (EDS) and Fourier-transform infrared spectroscopy (FTIR) on the Cu/GNS composite powders. To confirm the presence of GNS that reduced from GO, EDS analysis was performed. FTIR scanning is done on all the Cu/GNS composite powders to prove the presence of GNS.

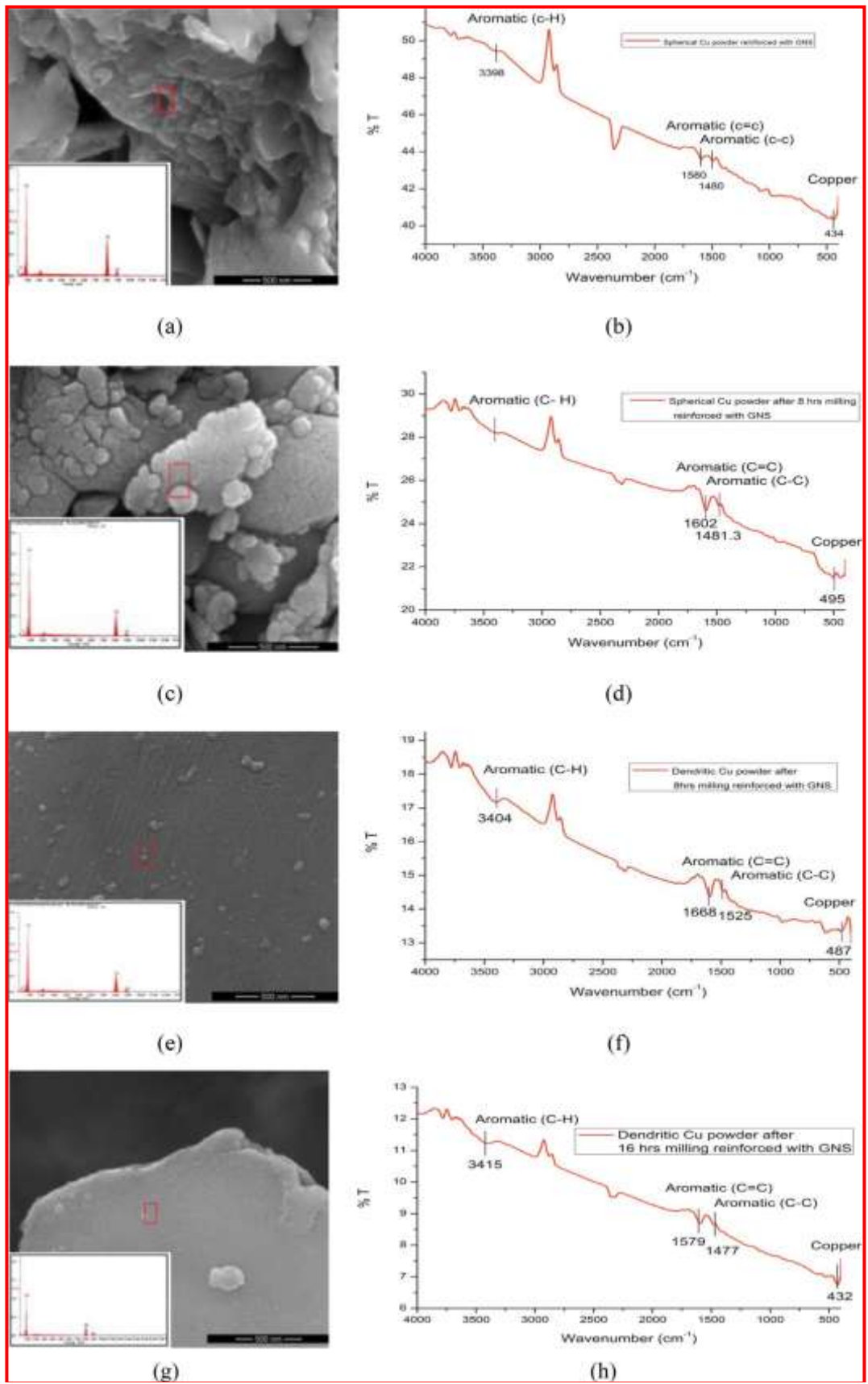

**Fig. 3.** Cu/GNS composite powders (a) SEM-EDS of spherical shape matrix Cu reinforced with GNS (b) FTIR of spherical shape matrix Cu reinforced with GNS (c) SEM-EDS of spherical Cu powder after 8 h milling reinforced with GNS (d) FTIR of spherical Cu powder

after 8 h milling reinforced with GNS (e) SEM-EDS of dendritic Cu powder after 8 h milling reinforced with GNS (f) FTIR of dendritic Cu powder after 8 h milling reinforced with GNS (g) SEM-EDS of dendritic Cu powder after 16 h milling reinforced with GNS (h) FTIR of dendritic Cu powder after 16 h milling reinforced with GNS.

Fig. 3(a–h) confirms Cu does not form carbides; the carbon peak in EDS is an evidence of GNS in the sample. Oxide peak is invisible in EDS analysis. Since the heating was done in inert gas atmosphere. All these eight images show a very good distribution of GNS in the Cu particles. The dendritic shape of the metal particles after 8 h milling with its higher specific surface area expected to enhance GNS distribution within the dendritic arms of the metal powders. The FTIR reveals that all the Cu/GNS composites have C=C (1579–1668), CeC (1477–1525) bands which indicates the presence of GNS [26–28]. From FTIR it is clearly seen that no peaks is observed in between 3000–3500 which shows the good reduction of GNS from graphene oxide. In addition to that no Cu oxide peaks is present in the Cu/GNS composites. PVA decomposition as carbon dioxide is clearly confirmed from the presence of carbon dioxide peaks in Cu/GNS composites.

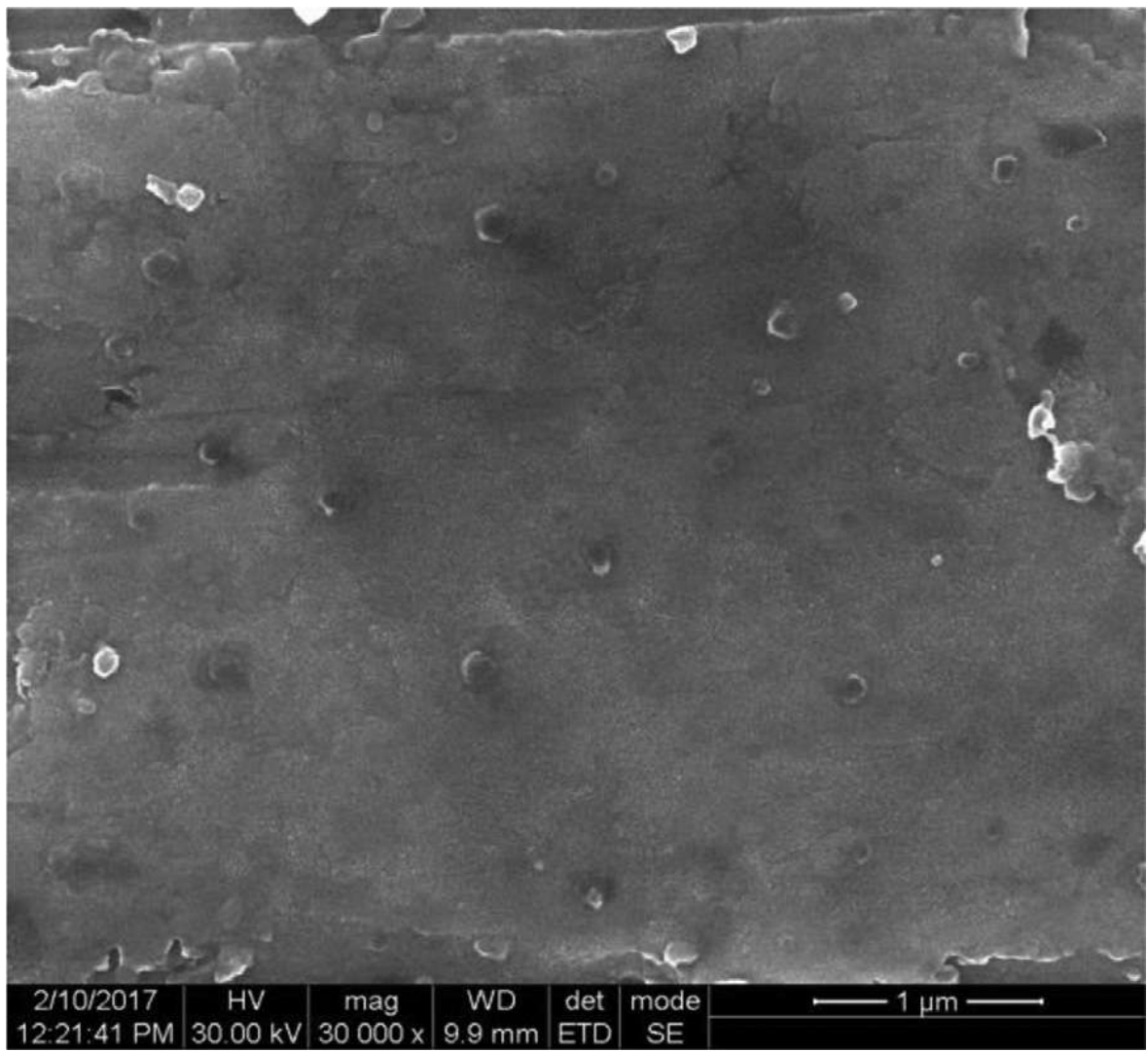

**Fig. 4.** Cu/GNS composite after sintering.

SEM and TEM characterization results indicated that a homogeneous dispersion and a good combination between graphene and Cu matrix, as well as the intact structure of graphene, which was beneficial to its strengthening effect [34]. No pore or crack in the interface between Cu matrix and graphene sheet indicates well interfacial bonding. As a result, the excellent hardness is obtained in the RGO-Cu composites [35, 36]. Fig. 4 shows the Cu/GNS after sintering from the image it clearly seen the dispersion of GNS uniformly over the entire area of Cu particles. Good bonding between Cu and GNS exist. It is complicated to discover the pores and gaps.

### 3.2. Hardness

The clean and close interface is in favor of the load and phonon transfer between Cu matrix and graphene. High mechanical and thermal conductivity properties are achieved only upon addition of 0.3 wt% graphene [37]. Mechanical milling is a simple method to prepare nanopowder where milling time plays very important role [38]. The milling time closely related to hardness of the composite and particles deformation. Increasing of milling time leads to reduction in particles size and improvement in hardness [29]. Accordingly, 16 h milling sample exhibits the highest hardness. It is due to the dispersion of GNS filler into Cu powder matrix which is achieved by the high energy ball milling. The interface between the matrix and filler supports for the loading of the filler into matrix. In turn, the loading offers a very good phonon transfer between the matrix and the filler [30].

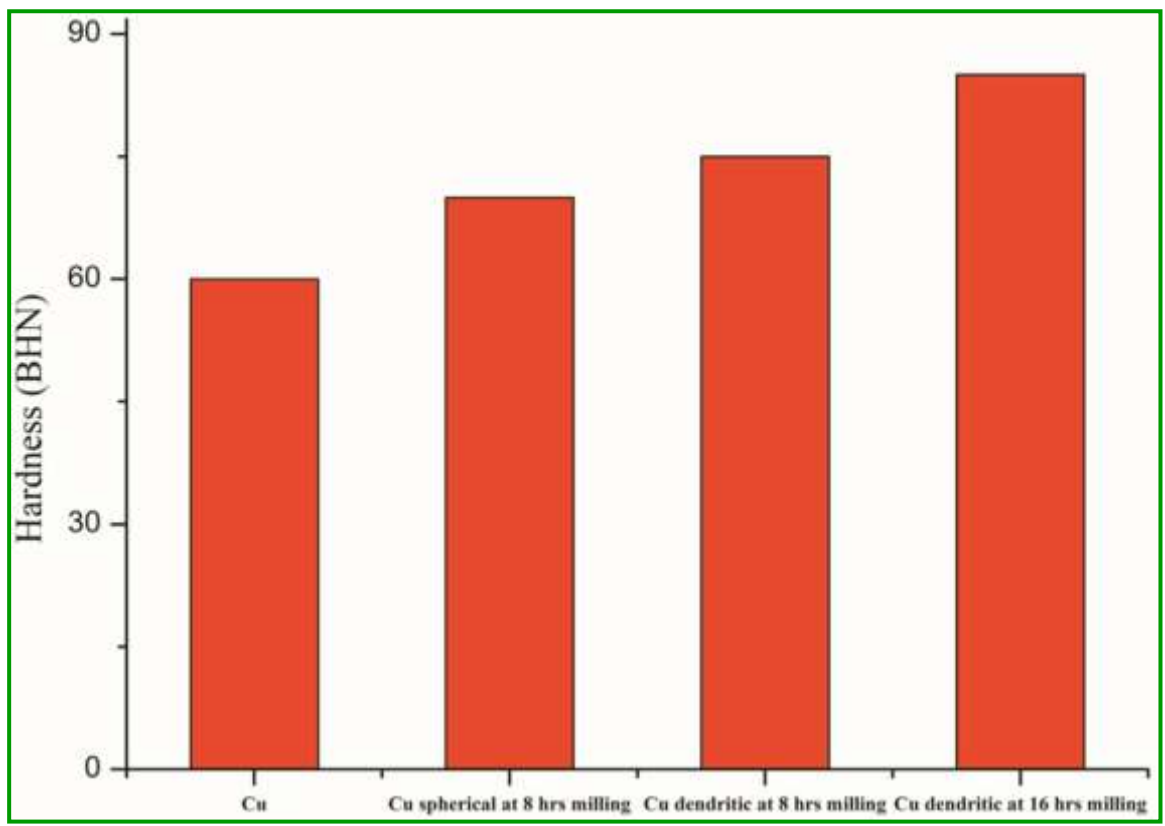

**Fig. 5.** Effect of hardness.

The hardness of Cu/GNS composites was measured in a Brinell scale (29 kN force and with a steel ball). Fig. 5 shows the Brinell Hardness No. (BHN) of Cu/GNS composites fabricated with 45 μm spherical Cu powder (60 BHN), spherical Cu powder after 8 h milling (70 BHN), dendritic Cu powder after 8 h milling (75 BHN), nanostructure flake Cu powder after 16 h milling (83 BHN). The hardness of composites fabricated with GNS 6 h milling, Hardness of Cu/ GNS (Cu after 16 h milling) composites is increased up to 20% by using flake Cu particles at low thickness. However, the best hardness value (81 BHN) was achieved in the composite fabricated with nanostructure dendritic Cu particles at 2 wt% GNS. The better hardness value attains at dendritic Cu powder milled for 16 h. The enhancement in hardness attained with the effect of grain refinement and strain hardening during milling. This effect is due to Hall–Petch relationship i.e. decreasing particles size increases hardness of the composite. The addition of reinforcement also has the same effect to enhance the hardness by uniform dispersion and interfacial strength. Uniform dispersion of reinforcement has ability to control the load transfer without uniformity in dispersion exhibits failure. Since all the prepared composites shows better hardness than pure Cu with consequence of even uniform of GNS.

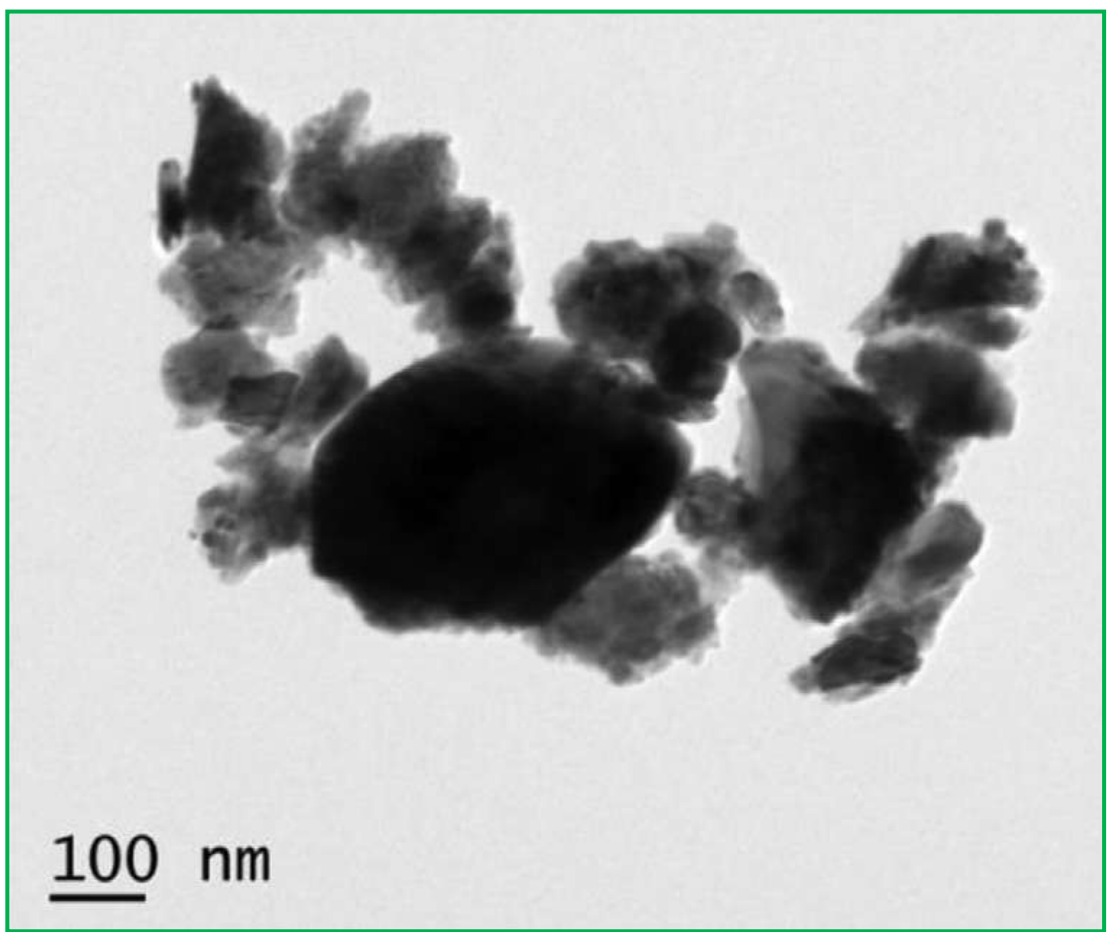

**Fig. 6.** TEM image of Cu/GNS composite

There are two effects for increasing strength of the Cu/GNS composite. First one is Hall–Petch relationship, with particle size reduction the hardness increased. Fig. 6 shows the grain boundary of GNS, grain boundaries of GNS acts as efficient constraints to disorder propagation across the Cu/GNS interface.

## 4. Conclusion

In this work, the effects of various morphology Cu powders were reinforced with GNS and their hardness investigated after sintering. GNS were successfully reinforced with various morphology Cu powders as matrix. It proves that powder metallurgy assisted with PVA treatment method is a promising one for preparing Cu/GNS composite. The uniform dispersion of GNS is seen in different morphology Cu powders. The hardness of the composites were evaluated with uniform reinforcement of 2 wt% GNS with various morphology Cu powders. It is found that all the prepared Cu/GNS composites shows better hardness than pure Cu. Ultimately Cu powders with dendritic shape after 16 h milling shows better hardness. It is 20% more than pure Cu. This hardness enhancement attained with the effect of uniform dispersion of GNS and also with good interfacial bonding. It is concluded that authors proposed method for preparing Cu/GNS composite is suitable for uniform dispersion of GNS with various morphology Cu powders.

**Acknowledgment**

Financial assistance from Board of research in nuclear science, Department of atomic energy, Govt. of India under the project sanction no.34/14/64/2014-BRNS/2140 is gratefully acknowledged.